\input harvmac
\noblackbox

\font\cmss=cmss10 \font\cmsss=cmss10 at 7pt
 \def\inbar{\,\vrule height1.5ex width.4pt depth0pt}
\def\IZ{\relax\ifmmode\mathchoice
{\hbox{\cmss Z\kern-.4em Z}}{\hbox{\cmss Z\kern-.4em Z}}
{\lower.9pt\hbox{\cmsss Z\kern-.4em Z}}
{\lower1.2pt\hbox{\cmsss Z\kern-.4em Z}}\else{\cmss Z\kern-.4em
Z}\fi}
\def\IB{\relax{\rm I\kern-.18em B}}
\def\IC{{\relax\hbox{$\inbar\kern-.3em{\rm C}$}}}
\def\ID{\relax{\rm I\kern-.18em D}}
\def\IE{\relax{\rm I\kern-.18em E}}
\def\IF{\relax{\rm I\kern-.18em F}}
\def\IG{\relax\hbox{$\inbar\kern-.3em{\rm G}$}}
\def\IGa{\relax\hbox{${\rm I}\kern-.18em\Gamma$}}
\def\IH{\relax{\rm I\kern-.18em H}}
\def\II{\relax{\rm I\kern-.18em I}}
\def\IK{\relax{\rm I\kern-.18em K}}
\def\IR{\relax{\rm I\kern-.18em R}}


\def\wall{$W(\chi^+,\chi^-)$}

\lref\senreview{For a review of this program 
with extensive 
references see: A. Sen, ``Non-BPS States and Branes in String Theory,''
hep-th/9904207.}
\lref\sennon{A. Sen, ``BPS D-Branes on Non-Supersymmetric Cycles,''
hep-th/9812031.}
\lref\Fayet{P. Fayet, ``Higgs Model and Supersymmetry,'' Nuovo Cim.
{\bf 31A} (1976) 626.}
\lref\vafa{C. Vafa, ``Extending Mirror Conjecture to
Calabi-Yau with Bundles,'' hep-th/9804131.}
\lref\joyce{D. Joyce, ``On counting special Lagrangian homology
3-spheres," hep-th/9907013.}  
\lref\bdl{M. Berkooz, M. Douglas, and R. Leigh, ``Branes Intersecting
at Angles," Nucl. Phys. {\bf B480} (1996) 265, hep-th/9606139.}
\lref\ibanez{L. Ibanez et al, D-terms in type IIB orientifolds}
\lref\ks{S. Kachru and E. Silverstein, ``4d Conformal Field Theories
and Strings on Orbifolds,'' Phys. Rev. Lett. {\bf 80} (1998) 4855,
hep-th/9802183\semi
S. Kachru, J. Kumar and E. Silverstein, ``Vacuum Energy Cancellation 
in a 
Nonsupersymmetric String,'' Phys. Rev. {\bf D59} (1999) 106004, hep-th/9807076.}
\lref\hm{J. Harvey and G. Moore, ``Algebras, BPS States, and 
Strings,'' Nucl. Phys. {\bf B463} (1996) 315, hep-th/9510182\semi
J. Harvey and G. Moore, ``On the Algebras of BPS States,'' Comm. Math.
Phys. {\bf 197} (1998) 489, hep-th/9609017\semi 
J. Harvey and G. Moore, ``Superpotentials and Membrane Instantons,''
hep-th/9907206.}
\lref\bbs{K. Becker, M. Becker and A. Strominger, ``Five-Branes,
Membranes and Nonperturbative String Theory,'' Nucl. Phys. {\bf B456}
(1995) 130, hep-th/9507158.}
\lref\bdlr{I. Brunner, M. Douglas, A. Lawrence and C. Romelsberger,
``D-branes on the Quintic,'' hep-th/9906200.}
\lref\klm{A. Karch, D. L\"ust and A. Miemiec, ``N=1 Supersymmetric 
Gauge Theories and
Supersymmetric Three Cycles," hep-th/9810254.}
\lref\syz{A. Strominger, S.T. Yau and E. Zaslow, ``Mirror Symmetry is 
T-Duality," Nucl. Phys. {\bf B479} (1996) 243, hep-th/9606040.}
\lref\ooy{H. Ooguri, Y. Oz and Z. Yin, ``D-Branes on Calabi-Yau
Spaces and their Mirrors,'' Nucl. Phys. {\bf B477} (1996) 407, 
hep-th/9606112\semi
K. Becker, M. Becker, D. Morrison, H. Ooguri, Y. Oz and Z. Yin, 
``Supersymmetric Cycles in Exceptional Holonomy Manifolds and
Calabi-Yau 4 Folds,'' Nucl. Phys. {\bf B480} (1996) 225, hep-th/9608116.} 
\lref\hitchin{N. Hitchin, ``The moduli space of special Lagrangian
submanifolds,'' math.dg/9711002\semi 
N. Hitchin, ``Lectures on Special Lagrangian Submanifolds,'' math.dg/9907034.}
\lref\bsv{
M. Bershadsky, V. Sadov and C. Vafa, ``D-Strings on D-Manifolds,"
Nucl. Phys. {\bf B463} (1996) 398, hep-th/9510225\semi
M. Bershadsky, V. Sadov and C. Vafa, ``D-Branes and Topological
Field Theories,'' Nucl. Phys. {\bf B463} (1996) 420, hep-th/9511222.}
\lref\mclean{R.C. McLean, {\it Deformations and moduli of calibrated
submanifolds}.  PhD thesis, Duke University, 1990.}

\Title{\vbox{\baselineskip12pt\hbox{hep-th/9908135}
\hbox{SU-ITP-99/41}
}}
{\vbox{\centerline{
Supersymmetric Three-cycles and}
\centerline{(Super)symmetry Breaking}}
}
\centerline{Shamit Kachru$^{1}$ and John McGreevy$^{2}$ 
}
\bigskip
\bigskip
\centerline{$^{1}$Department of Physics and SLAC}
\centerline{Stanford University}
\centerline{Stanford, CA 94305}
\centerline{Email: skachru@leland.stanford.edu}
\smallskip
\medskip
\centerline{$^{2}$ Department of Physics}
\centerline{University of California at Berkeley}
\centerline{Berkeley, CA 94720}
\centerline{Email: mcgreevy@socrates.berkeley.edu}
\medskip
\bigskip
\medskip
\noindent
We describe physical phenomena associated with a class of transitions
that occur in the study of supersymmetric three-cycles 
in Calabi-Yau threefolds.  The transitions in question occur 
at real codimension one in the complex structure moduli space of
the Calabi-Yau manifold.
In type IIB string theory, these transitions can be used to describe the
evolution of a BPS state as one moves through a locus of marginal
stability:  
at the transition point the BPS particle becomes degenerate with a 
supersymmetric two particle state,
and after the transition the lowest energy state carrying the
same charges is a non-supersymmetric two particle state.
In the IIA theory, wrapping the cycles in question with D6-branes leads
to a simple realization of the Fayet model: for some values of the CY modulus
gauge symmetry is spontaneously broken, while for other values supersymmetry
is spontaneously broken. 

\Date{August 1999}

\newsec{Introduction}

In the study of string compactifications on manifolds of reduced holonomy,
odd-dimensional supersymmetric cycles 
play an important part
(see for instance
\refs{\bbs,\hm,\bsv,\ooy,\syz,\hitchin, 
\vafa,\klm,\bdlr,\joyce} and references therein).  
In type IIB string theory, a supersymmetric three-cycle can be
wrapped by a D3-brane to yield a BPS state whose properties 
are amenable to exact study; 
in the IIA theory or in M theory, Euclidean membranes can wrap
the three-cycle
and contribute to ``holomorphic'' terms in the 
low energy effective action of the spacetime
theory (that is, terms that are integrated over only a subset of the
fermionic superspace coordinates).

Of particular interest, partially 
due to their role in mirror symmetry \refs{\syz,\vafa}, 
have been 
special Lagrangian submanifolds
in Calabi-Yau threefolds.
In an interesting recent paper by Joyce \joyce, various transitions which these
cycles undergo as one moves in the complex structure or K\"ahler moduli
space of the underlying CY manifold were described. 
In this note, we study some of the physics associated with the
simplest such transitions discussed in \S6\ and \S7\ of \joyce.  
These transitions
are reviewed in \S2.  
The physical picture which one obtains by wrapping D3-branes on the
relevant cycles in IIB string theory is described in \S3, while the
physics of wrapped D6-branes in type IIA string theory occupies \S4.
Our discussion is purely local (in both the moduli space and the
Calabi-Yau manifold), as was the analysis performed in
\joyce; we close with some speculations about more
global aspects in \S5.

At all points in this paper, we will be concerned with $\it rigid$ special
Lagrangian three cycles.  Since the moduli space of a special Lagrangian
three cycle $N$ (including Wilson lines of a wrapped D-brane) is
a complex K\"ahler manifold of dimension $b_1(N)$ \refs{\mclean,\hitchin},
this means we have to focus on 
so-called
``rational homology three spheres'' with 
$H_1(N,\IZ)$ at most a discrete group.  We will further assume that 
$H_1(N,\IZ)$ is trivial.

\newsec{Splitting Supersymmetric Cycles}

\subsec{Definitions}

Let $M$ be a Calabi-Yau threefold equipped with a choice of complex
structure and K\"ahler structure. 
Let $\omega$ be the K\"ahler form on $M$, and let $\Omega$ be the
holomorphic three-form, normalized to satisfy
\eqn\norm{{\omega^{3}\over 3!} ~=~{i\over 8} \Omega \wedge \overline{\Omega}} 
This also allows us to define two real, closed three forms on $M$,
Re($\Omega$) and Im($\Omega$).

Let $N$ be an oriented real three-dimensional submanifold of $M$.
We call $N$ special Lagrangian with phase
$e^{i\theta}$ iff

\noindent
a)~ $\omega \vert_N ~=~0$ 

\noindent
b)~ $(sin(\theta) {\rm Re (\Omega)} - 
cos(\theta) {\rm Im (\Omega))} \vert_N ~=~0$

\noindent
(a) and (b) together imply that 

\eqn\svol{\int_N (cos(\theta) {\rm Re (\Omega)} + 
sin(\theta) {\rm Im (\Omega})) ~=~ vol(N)} 
where $vol(N)$ is the volume of $N$.

Physically, the relevance of $\theta$ for us will be the following.
Let $N$ and $N^\prime$ be three-cycles which are special Lagrangian with
different phases $\theta$ and $\theta^\prime$.  Compactifying, say, 
IIB string
theory on $M$, we can obtain BPS states which preserve
half of the ${\cal N}=2$ spacetime supersymmetry by wrapping three-branes
on $N$ or $N^\prime$.  In the notation of \bbs, the surviving supersymmetries 
in the presence of a D3-brane on $N$, for example,  are generated by 
$$ \epsilon_{\delta} = e^{i\delta}\epsilon_{+} + e^{-i\delta}\epsilon_{-}, $$
with $\delta = -{\theta \over 2} - {\pi \over 4}.$
For generic $\theta \neq \theta^\prime$, however,
$N$ and $N^\prime$ preserve different ${\cal N}=1$ supersymmetries and 
the state with both wrapped three-branes would 
break all of the supersymmetry. 

\subsec{Transitions}

The following supersymmetric three-cycle 
transitions are conjectured by Joyce to occur in
compact Calabi-Yau threefolds $M$.    

Choose two
homology classes 
$\chi^{\pm} \in H_{3}(M,\IZ)$ which are linearly independent in
$H_3(M,\IR)$.  
For any $\Phi \in H^{3}(M,\IC)$, define
\eqn\phidef{\Phi \cdot \chi^\pm ~=~\int_{\chi^\pm} \Phi}
Thus $\Phi \cdot \chi^\pm$ are complex numbers.    
Following Joyce, define a subset $W(\chi^+,\chi^-)$
in $H^{3}(M,\IC)$ by

\eqn\wdef{W(\chi^+,\chi^-)~=~\{ \Phi \in H^{3}(M,\IC) : 
(\Phi \cdot \chi^+) (\overline \Phi \cdot \chi^-) \in (0,\infty) \}}

\noindent
So $W(\chi^+,\chi^-)$ is a real hypersurface in $H^3(M,\IC)$.

Fix some small, positive angle $\epsilon$.  For $\Phi \in H^{3}(M,\IC)$
write
$$(\Phi \cdot \chi^+) (\overline \Phi \cdot \chi^-) ~=~R e^{i\theta}$$
where $R \geq 0$ and $\theta \in (-\pi,\pi]$. Then we say $\Phi$ lies
in $W(\chi^+,\chi^-)$ if $R > 0$ and $\theta = 0$. 
We say that $\Phi$ lies on the positive side of $W(\chi^+,\chi^-)$ if
$R > 0$ and $0 < \theta < \epsilon$.  We say that $\Phi$ lies on the
negative side of $W(\chi^+,\chi^-)$ if $R > 0$ and $-\epsilon < \theta
< 0$.  
Then, Joyce argues that the following kinds of transitions should occur.
We are given a Calabi-Yau $M$ with compact, nonsingular 
three cycles $N^\pm$ in homology classes 
$[N^\pm] = \chi^\pm$.  $N^\pm$ are taken to be 
special Lagrangian with phases $\theta^\pm$.
We assume $N^\pm$ intersect at one point $p \in M$, with $N^+ \cap 
N^-$ a positive intersection.  As we deform the complex structure of $M$,
the holomorphic three form moves around in $H^3(M,\IC)$ and
therefore the phases $\theta^\pm$ of $N^\pm$ change.

When $[\Omega]$ is on the positive side of $W(\chi^+,\chi^-)$ 
there exists a special Lagrangian threefold $N$ which is diffeomorphic
to the connected sum $N^+ \# N^-$, with $[N] = [N^+] + [N^-]$
in $H_3(M,\IZ)$.  
$N$ can be taken to be special Lagrangian with phase $\theta = 0$ (this
fixes the phase of $\Omega$ for us).
As we deform $[\Omega]$ through $W(\chi^+,\chi^-)$,
$N$ converges to the singular union $N^+ \cup N^-$. 
When $[\Omega]$ is in $W(\chi^+,\chi^-)$, the phases $\theta^\pm$ align
with $\theta = 0$.
On the negative 
side of $W(\chi^+,\chi^-)$, $N$ ceases to exist as a special 
Lagrangian submanifold of $M$ (while $\theta^\pm$ 
again become distinct). 

For completeness and to establish some notation we will find 
useful, we briefly mention some motivation for the existence 
of these transitions \joyce.
In Joyce's model of the transition, there exists a 
manifold, $D$, with boundary $S \subset N$, which 
is special Lagrangian with phase $i$. If we call its volume $A$, this
means that $i A = \int_D \Omega$.  
$S$ defines a 2-chain in $N$;  
since we are assuming that $H_1(N,\IZ)$ 
is trivial, 
by Poincar\' e duality,
$S$ must be trivial in homology.  Because $S$ is real codimension one in $N$,
it actually splits $N$ into two parts:
$$N = C^+ \cup C^-, ~\del C^+ = -S,~ \del C^- = S.$$
So $C^{\pm} \pm D$ define 3-chains and in fact it turns out that 
\eqn\nd{[C^{\pm} \pm D] = \chi^\pm = [N^\pm]}
We see that we can determine the volume of $D$ just from knowledge 
of $\chi^\pm$:
\eqn\Afromchi{  A = {1 \over i} \int_D \Omega = \int_D {\rm Im (\Omega)} = 
\int_{\chi^+} {\rm Im (\Omega)} }
using ${\rm Re (\Omega)} |_D = 0$ and ${\rm Im (\Omega)} |_N = 0$.  
But when $[\Omega]$ goes through \wall, we see from \Afromchi\ 
and from the definition of \wall\ 
that $A$ becomes negative; at least in the local model in $\IC^3$, this 
means that $N$ does not exist.

\newsec{Formerly BPS States in IIB String Theory}

Now, consider Type IIB string theory compactified on $M$. 
When the complex structure is such that $[\Omega]$ is on the positive
side of $W(\chi^+,\chi^-)$, one can obtain a BPS hypermultiplet by
wrapping a D3-brane on $N$.  One can also obtain BPS hypermultiplets
by wrapping D3-branes on $N^+$ or $N^-$.  

Because
$$[N] ~=~[N^+] + [N^-]$$
one can make a state carrying the same charges as the BPS brane wrapping
$N$ by considering the two particle state with D3-branes wrapping
both $N^+$ and $N^-$.   
How does the energy of the two states compare?

Recall that the disc $D$ with boundary on $N$ splits $N$ into two components,
$C^\pm$.  Define
\eqn\bdef{B^\pm ~=~\int_{C^\pm} \Omega}
Then if we let $V$ denote the volume of $N$ and $V^\pm$ denote the volumes
of $N^\pm$, we recall:
\eqn\v{V~=~B^+ + B^-}
\eqn\vpm{V^{\pm} e^{i\theta^\pm} = B^{\pm} \pm iA}
where $A$ is the volume of $D$.  
Since on this side of the transition $A$ is positive, $\theta^+$ is small and
positive while $\theta^-$ is small and negative.
In fact, reality of the volumes $V^\pm$ lets us solve for $\theta^\pm$
in terms of $B^\pm$ yielding
\eqn\thetapm{\theta^\pm ~=~\pm {A\over B^\pm}}

The energy of the single particle state obtained by wrapping a D3-brane
on $N$ is $T_{D3} \times V$ where $T_{D3}$ is the
D3 brane tension.  The energy of the (nonsupersymmetric) 
state obtained by wrapping
D3-branes on both $N^\pm$
can be approximated by $T_{D3} \times (V^+ + V^-)$.  
Expanding \vpm\ for small $\theta^\pm$, we find:
\eqn\bound{V^+ + V^- ~=~V + A ~(\theta^+ - \theta^-)~=~V + 
A^{2}~({1\over B^+} + {1\over B^-}) }
So 
since $A > 0$ and $\pm \theta^\pm > 0$ on this side of the
transition, we see that the single wrapped brane on $N$ is
energetically preferred.

Therefore, when the complex structure is on the positive side of
$W(\chi^+,\chi^-)$, the BPS state indeed has lower energy than the
nonsupersymmetric two particle state carrying the same charges, by
roughly $T_{D3}\times A (\theta^+ - \theta^-)$.

Now as one moves in the complex structure moduli space of $M$ through
a point where $[\Omega]$ lies in $W(\chi^+,\chi^-)$, 
$A$ and $\theta^\pm$ vanish.  Therefore, \bound\ shows that
that mass of the two particle state becomes equal to that of the
single particle state: we are passing through a locus of marginal
stability.  On this locus, the two particle state consisting of
branes wrapping $N^\pm$ is supersymmetric, since $N^\pm$ are
special Lagrangian with the same phase. 

Finally, move through to the region where $[\Omega]$ lies on the negative
side of $W(\chi^+,\chi^-)$. 
Here, $\pm \theta^\pm < 0$.  Since $N$ ceases to exist as a
supersymmetric cycle, the two particle
state with D3-branes wrapping $N^\pm$ is the lowest energy state
carrying its charges.\foot{In a global model, even if 
there do exist other supersymmetric 
cycles in the same class, there will be some 
region in moduli space close to the transition where 
the energy cost for moving to them in the 
Calabi-Yau will be larger than the energy gained.}  
Note that the two particle state is $\it nonsupersymmetric$, since
$N^\pm$ are special Lagrangian with different phases. 
Here, we are
making the conservative assumption that there is no stable, nonsupersymmetric
bound state of these two particles -- such a bound state would be
reflected in the existence of a (nonsupersymmetric) cycle in the homology
class $[N^+] + [N^-]$ with lower volume than $V^+ + V^-$.
This is tantamount to
assuming that the force between the two particles is repulsive for slightly 
negative $A$.  This is reasonable since 
for $A$ positive there is an attractive force and a (supersymmetric)
bound state, and as $A$ decreases to zero the magnitude of the
force and the binding energy 
decrease
until they vanish when $A=0$. 

This phenomenon is an interesting variant on the examples of \sennon.
There, a stable nonsupersymmetric state 
passes through a locus of marginal
stability and 
becomes unstable to 
decay to a pair of BPS particles (which together 
break all of the supersymmetries).
In the present example, 
a BPS particle becomes, as we move in complex structure  
moduli space, unstable to decay to a pair of BPS particles.  Moving
slightly further in moduli space, we see that the two BPS particles together
break all of the supersymmetries.

\newsec{D6-Branes and the Fayet Model} 

Now, consider type IIA string theory on the Calabi-Yau $M$ in which the
phenomena of \S2\ are taking place.  
Instead of studying particles in the resulting ${\cal N}=2$ supersymmetric
theory, we 
wrap the three-cycle $N$ with a space-filling
D6-brane (i.e., 3+1 of the dimensions of the D6-brane fill the non-compact
space).  This yields an ${\cal N}=1$ supersymmetric theory
in the non-compact dimensions.  For 
simplicity (since all our considerations are local), we 
can assume $M$ is non-compact so we do not have to worry about cancelling
the D6 Ramond-Ramond charge.  
Alternatively, we could imagine the model discussed below arising as part
of a larger system of branes and/or orientifolds on $M$. 

First, let's discuss the physics when $[\Omega]$ is on the positive side of
$W(\chi^+,\chi^-)$.
Since $b_1(N) = 0$, $N$ has no moduli in $M$.  Therefore,
there are no moduli in the effective 3+1 dimensional field theory
on the wrapped D6-brane.  The $U(1)$ gauge field on the brane
survives reduction on $N$, so the 3+1 dimensional 
low energy effective theory has a $U(1)$ gauge symmetry. 
Finally, because $N$ is a supersymmetric cycle
with $H_{1}(N,\IZ)$ trivial,
there is a unique supersymmetric ground state in the gauge
theory (as opposed to a discrete set of ground states parametrized by
Wilson lines around $N$). 

What about the physics when $[\Omega]$ is on the negative side of
$W(\chi^+,\chi^-)$?  The D6 which was wrapping $N$ has now split into
two D6-branes, wrapping $N^+$ and $N^-$.  
The $U(1)$ gauge field on each survives, yielding a $U(1)^2$ gauge theory.
Because $N^+$ and $N^-$ are
supersymmetric cycles with different phases, the theory
has no supersymmetric ground state.  We do expect
a stable nonsupersymmetric ground state, as long as $[\Omega]$ is close enough
to $W(\chi^+,\chi^-)$.   

What is the physics associated with the phase transition when $[\Omega]$
lies in $W(\chi^+,\chi^-)$?  At this point, the two D6-branes wrapping
$N^+$ and $N^-$ preserve the same supersymmetry, and intersect at a 
point in $M$. 
Because the light states
are localized at the intersection, the global geometry of the intersecting 
cycles doesn't matter and we can model the physics by a pair of 
flat special Lagrangian three-planes intersecting at a point. 
This 
kind of 
system was discussed in \bdl, and using their results 
it is easy to see that the resulting light strings give 
rise to precisely one chiral multiplet with charges $(+,-)$ under the
$U(1)^2$ gauge group of the two wrapped D branes.  Therefore,
one linear combination
of the $U(1)$s (the normal ``center of mass'' $U(1)$) remains 
free of charged matter, while the other (the ``relative" $U(1)$) gains
a single charged chiral multiplet $\Phi$.
The relative $U(1)$ is therefore anomalous; \bdl\ demonstrates that the 
anomaly is cancelled by inflow from the bulk.

Ignoring the center of mass $U(1)$ (which we identify with the surviving
$U(1)$ on the positive side of $W$), the physics of this model is 
precisely reproduced by 
the Fayet model, the simplest model of
spontaneous (super)symmetry breaking \Fayet.
This is a $U(1)$ gauge theory with a single charged chiral multiplet 
$\Phi$ (containing a complex scalar $\phi$).
There is no superpotential, but including a Fayet-Iliopoulos term $rD$
in the spacetime Lagrangian, the potential energy is 
\eqn\fpot{V(\phi) ~=~{1\over g^2}~ (\vert \phi \vert^{2} - r)^{2}}
where $g$ is the gauge coupling.

The phase structure of the model is quite simple: For $r>0$, there is 
a unique supersymmetric minimum, 
and the $U(1)$ gauge symmetry is Higgsed.  For $r<0$, there is a
unique nonsupersymmetric minimum at $\phi=0$, so the $U(1)$ symmetry
is unbroken.  Precisely when $r=0$, there 
is a $U(1)$ gauge theory with a massless charged chiral field and
a supersymmetric ground state.  

Thus, we are led to identify the regions of positive, vanishing and
negative $r$ with the positive side of $W(\chi^+,\chi^-)$, the locus
where $[\Omega]$ is in $W$, and the negative side of $W$.
The single real modulus which varies in the transition 
experienced by the supersymmetric three-cycle $N$ 
can be identified with the Fayet-Iliopoulos
parameter $r$.  
This identification is 
consistent with the conjecture in \bdlr\ that in worldvolume 
gauge 
theories of A-type D-branes on Calabi-Yau spaces, complex structure moduli 
only enter as D-terms.\foot {Note that the D6 branes in question here are 
considered A-type branes in the conventions of \bdlr\ since the 
three non-compact spatial dimensions are ignored.}

\newsec{Discussion}

Exploration of the 
phenomena involving supersymmetric cycles in a Calabi-Yau
manifold $M$ under variation of the moduli of $M$ has just started. 
It should be clear that as such phenomena are understood, 
they will have interesting implications for the physics of D-branes on
Calabi-Yau spaces (for a nice discussion of various aspects of this,
see \bdlr). 

One of the most enticing possibilities is that as more such phenomena
are uncovered, we will find new ways to
``geometrize" the study of supersymmetry breaking models in string theory.
This would provide a complementary approach to attempts to write down
interesting nonsupersymmetric string models informed by AdS/CFT
considerations \ks\ 
or insights about tachyon condensation and nonsupersymmetric
branes \senreview.

As a small step in this direction, it would be nice to find ways of going
over small potential hills between different supersymmetric vacua of 
string theory.  The transitions studied here, when put in the more global
context of a manifold $M$ with (possibly) several supersymmetric cycles in
each homology class, might provide a way of doing this.  
For instance in \S4, as one moves $[\Omega]$ into the negative side of
$W(\chi^+,\chi^-)$, it is clear that one is increasing the scale
of supersymmetry breaking (at least in the region close to the transition).
Suppose that after one moves through the negative side of $W$ in
complex structure moduli space, eventually $N^+$ and $N^-$ approach each
other and intersect again 
and the phenomenon of \S2\ occurs in reverse, with a new supersymmetric
cycle 
$N^\prime$ in the same homology class as $[N^+] + [N^-]$ popping into
existence. 
In such a case, one would have a nonsupersymmetric ground state for some range
of parameters on the negative side of $W$, and then eventually reach
a supersymmetric ground state again (with the D6 brane wrapping
$N^\prime$).  

Similarly, on the negative side of $W$ there could exist
``elsewhere" in $M$ a supersymmetric cycle $\tilde N$ in the same class as
$[N^+] + [N^-]$.  Although the cost in energy to move from wrapping
$N$ to wrapping $\tilde N$ is nonzero and hence on the negative side
of $W$ the phenomena of \S3, \S4 occur, eventually it may become
advantageous for the D6 branes to shift over to wrapping $\tilde N$.
This would again be a situation where supersymmetry is
broken, and then restored, as one dials the complex structure modulus
of the Calabi-Yau space.

\centerline{\bf{Acknowledgements}}

We are grateful to J. Harvey, G. Moore and E. Silverstein for discussions.
The research of S.K. is supported by an A.P. Sloan Foundation Fellowship
and a DOE OJI Award.  The research of J.M. is supported by
the Department of Defense NDSEG Fellowship program.

\listrefs
\end